\documentclass[runningheads]{llncs}

\usepackage[T1]{fontenc}
\usepackage[utf8]{inputenc}
\usepackage{graphicx}
\makeatletter
\define@key{Gin}{alt}{}
\makeatother
\usepackage{hyperref}
\usepackage{url}
\usepackage{amsmath}
\usepackage{amssymb}
\usepackage{booktabs}
\usepackage{array}
\usepackage{caption}
\usepackage{wrapfig}
\usepackage{tikz}
\usetikzlibrary{arrows.meta,positioning}
\newcommand{\up}{\hat{u}}
\newcommand{\lft}{\hat{\ell}}

\begin{document}

\title{Rhythmic Fitness Beat 'Em All using Real-Time Monocular Human Pose Estimation}

\titlerunning{Rhythmic Fitness Beat 'Em All}

\author{Yanis Zaoui\inst{1,2} \and Valentin B\'egel\inst{2} \and Mohamed Adjel\inst{1}}

\authorrunning{Y. Zaoui et al.}

\institute{NaturalPad, Montpellier, France \and Universit\'e Paris Cit\'e, I3SP, URP 3625, Paris, France}

\maketitle

\begin{abstract}
Short and frequent physical activity exercices mitigate the health risks of sedentary behavior. Yet, the lack of motivation remains a critical barrier to adoption. Exergames gamify physical activity, but most rely on dedicated hardware, and outside dance and rhythm games they rarely use the musical beat. We present a work-in-progress beat 'em up exergame played with the whole body in front of a single smartphone, which runs the entire pipeline: an off-the-shelf real-time 3D human mesh recovery model on the phone's neural processing unit, gesture recognition, and the game, shown on any TV through a USB-C to HDMI cable. Our contributions are (1) a 3D recognition of five shadow-boxing moves (guard, punch, slip, squat, jump), each defined by one metric measure in a body frame aligned with gravity and one threshold with hysteresis, personalised during a tutorial; (2) a rhythm stage whose discrete cues are placed on the song's kick drum and judged on the movement's capture time, with a delay learned for each player; and (3) a phone-only architecture, running at 30 frames per second, that turns a phone and an HDMI cable into an exergame console.
\end{abstract}

\section{Introduction}

Physical inactivity is widespread and rising: in 2022, 31\% of adults worldwide did not meet the recommended levels of physical activity~\cite{Strain2024}, a trend increased by the global shift toward remote working~\cite{Schone2025}. Mitigating its health risks does not necessarily require lengthy, exhausting workouts. Research into ``exercise snacks'' shows that frequent and brief bouts of physical activity reduce the negative metabolic effects of prolonged sitting~\cite{Fang2026}, and as little as three bursts of one to two minutes of vigorous activity a day is associated with a mortality about 40\% lower~\cite{Stamatakis2022}. Despite the proven efficiency of these micro-workouts, the primary barrier to adoption remains the lack of motivation: physical activity is perceived as a chore rather than a source of pleasure. One of the most promising ways to ``hack'' this motivational deficit is by gamifying physical activity~\cite{Monroe2023}, transforming routine movements into engaging experiences~\cite{Grech2024}.

In this context, exergames have emerged as a dedicated category of physically interactive games. Following Oh and Yang~\cite{OhYang2010}, we define exergames as video games requiring physical exertion, more than sedentary activity, that may also include strength, balance, and flexibility components. Exergames reach light-to-moderate intensity~\cite{Peng2011}, and up to vigorous intensity for VR rhythm boxing~\cite{Craig2024}. However, the real-time motion capture they rely on usually requires dedicated hardware (motion controllers, depth cameras, VR headsets), whose cost, setup and space requirements make them less accessible, in particular for short sessions at home or at work. Moreover, although moving to music improves affect and performance during exercise~\cite{Terry2020}, rhythm remains a minority design element outside dance and rhythm games: only 10 of 186 papers on extended reality exergames use rhythmic tasks~\cite{Karaosmanoglu2024}.

Our goal is an exergame that needs nothing more than a smartphone and a TV, and that uses the musical beat as a central game mechanic. We built it on \textit{City of Rage}~\cite{CityOfRage}, a fan-made tribute to the arcade beat 'em up \textit{Streets of Rage 2}~\cite{SOR2}, whose vocabulary (punches, dodges, guard) is that of shadow boxing. The phone runs a real-time 3D human mesh recovery model, InstantHMR~\cite{InstantHMR}, which we use off the shelf, recognises the player's moves in 3D, and runs the game, displayed on a TV over a USB-C to HDMI cable. Our contributions are: (1) a 3D recognition method for five shadow-boxing moves, each defined by one metric measure in a calibrated body frame and one threshold, with re-arming and personal thresholds; (2) a rhythm stage with discrete cues placed on the song's kick drum and judged on the movement's capture time, with a delay learned for each player; and (3) a phone-only architecture that turns a phone and an HDMI cable into an exergame console.

\section{Related Work}
\label{sec:related}

\textbf{Exergames and discrete interaction.} Early exergames such as Dance Dance Revolution and Nintendo Wii Sports established an interaction paradigm that still dominates the field: mapping discrete gesture recognition to discrete game events. Discrete event detection is computationally tractable and unlocks reactive game mechanics. However, treating body movements as buttons increases the complexity of interaction as the number of mapped gestures grows, often leading to oversimplified gameplay constrained by a limited set of discrete commands. Recent webcam-based exergames have shown that computer vision can replace dedicated sensors, but few are thoroughly described or technologically robust~\cite{Bayd2025}.

\textbf{Music, rhythm, and movement.} Music is a powerful leverage for exercise commitment. Listening to music during physical activity increases motivation, improves mood, reduces perceived exertion, and enhances overall exercise performance~\cite{Ballmann2021,Terry2020}. Humans are remarkably good at picking up periodic signals: when a metronome ticks, we spontaneously tap along with a precision measured in tens of milliseconds~\cite{Repp2005,Repp2013}. This capacity has been leveraged in clinical rehabilitation, where rhythmic auditory cueing improves gait in populations with Parkinson's disease~\cite{Thaut2019,Wang2022} and post-stroke patients~\cite{BraunJanzen2021}. The privileged link between auditory rhythm and the motor system, which drives spontaneous entrainment and anticipatory movement alignment, has been extensively characterised~\cite{Whitton2023,Pranjic2024,Grahn2009,Large1999}. Music is also widely used in fitness and by individual sports practitioners~\cite{Karageorghis2012}.

\textbf{Music-driven and movement-adaptive games.} Successful music-based games such as Beat Saber demonstrated that coupling motion game mechanics with music is a compelling direction. Researchers have gone further, adapting music to movement in real time: BeatWalk uses an ankle IMU band to adapt musical tempo to the gait cadence of Parkinson's patients~\cite{CochenDeCock2021}, and serious games such as Rhythm Workers have demonstrated benefits on rhythmic motor skills in clinical populations~\cite{Begel2017,Begel2018,Puyjarinet2022}. These approaches, however, either target only fine movements (e.g., tactile finger tapping for Rhythm Workers) or require specific hardware (a VR headset for Beat Saber, an IMU for BeatWalk).

\textbf{Monocular pose estimation.} To remove this hardware dependency, one can leverage real-time monocular human pose estimation, which estimates the 2D or 3D position of human keypoints (e.g., joints) from a single camera~\cite{mediapipe,Sarandi2024}. Traditionally computationally heavy and reliant on dedicated GPUs~\cite{openpose}, this task can now run in real time on laptops and smartphones thanks to efficient architectures~\cite{mediapipe}, and smartphone exergames with markerless motion capture are an active topic~\cite{Phosanarack2025,Zhao2025,Richard2025}. Most of them rely on 2D keypoints, whose image distances depend on the player's distance and orientation to the camera. Human mesh recovery goes further and estimates the full 3D body, as the pose and shape parameters of a parametric body model~\cite{Kanazawa2018}; recent foundation models such as SAM 3D Body~\cite{Yang2026SAM3DBody}, built on the Momentum Human Rig (MHR)~\cite{Ferguson2025MHR}, are robust in the wild but far too heavy for a phone. We use InstantHMR~\cite{InstantHMR}, a lightweight MHR-based model that runs in real time on a smartphone's neural processing unit, so that every gesture threshold can be expressed in metres and degrees.

\section{System}
\label{sec:methods}

\begin{figure}[t]
  \centering
  \includegraphics[width=\textwidth,alt={System overview. Left: a phone shows the player throwing a punch, filmed by its camera at 30 fps. An arrow leads to a box: InstantHMR on the NPU, 3D body mesh. An arrow leads to the same frame with the 3D skeleton drawn over the player, labelled 3D Gestures: punch, slip, squat, jump, guard. An arrow leads to a TV showing City of Rage, where the fighter Axel punches the boss Curls, labelled cues on the kicks, any TV, USB-C to HDMI. Footer: everything runs on the phone, no PC, about 5.6 W.}]{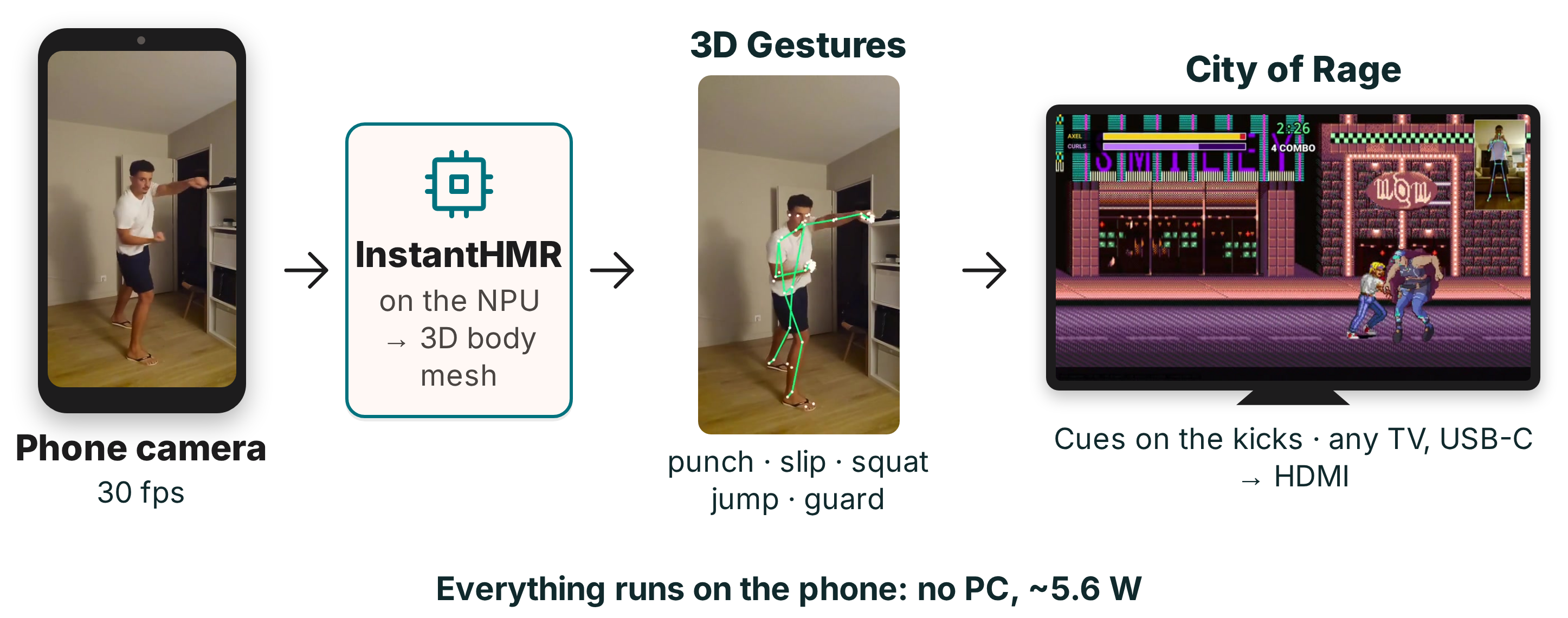}
  \caption{System overview. The phone's camera feeds InstantHMR~\cite{InstantHMR} on the phone's neural processing unit; the game recognises five moves on the resulting 3D skeleton and runs on the same phone, shown on a TV through a USB-C to HDMI cable. Everything runs on the phone, without a PC.}
  \label{fig:overview}
\end{figure}

We selected an open-source Pygame game, \textit{City of Rage}~\cite{CityOfRage}, a tribute to \textit{Streets of Rage 2}~\cite{SOR2}, and ported it to a web stack (JavaScript modules, Canvas 2D at a native resolution of $640\times360$, fixed 60~Hz simulation steps). Figure~\ref{fig:overview} gives an overview of the system. Everything runs on a single smartphone (a Samsung Galaxy S23 in our tests): a native Android app captures the front camera at 30 frames per second, finds the player with a NanoDet person detector~\cite{NanoDet}, and runs InstantHMR~\cite{InstantHMR} on the phone's neural processing unit (fp16), which outputs the player's 3D body mesh and 70 MHR keypoints in camera space, in metres. For each frame, the app streams the unsmoothed 3D skeleton and the frame's camera capture time over a local WebSocket to the web game, which runs in a WebView of the same app and is shown on a TV through a USB-C to HDMI cable. Gestures are recognised in the game, not in the pose source, so that any pose source streaming the same skeleton (e.g., a desktop webcam during development) drives the same game.

On the Galaxy S23, the pipeline runs at 30 frames per second (about 25 when the phone is hot), with 72--76~ms from camera capture to sending the skeleton (about 45~ms of it inside the phone's camera software), and the phone draws about 5.6~W while playing.

\subsection{Gesture Recognition in 3D}
\label{sec:gestures}

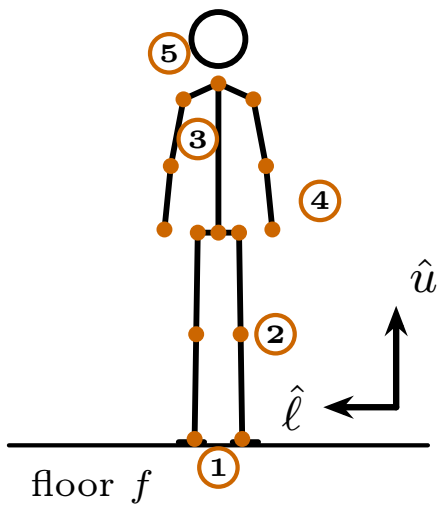
\begin{wrapfigure}{r}{0.36\textwidth}
  \vspace{-1.2\baselineskip}
  \centering
  \begin{tikzpicture}[x=0.0105cm, y=-0.0105cm, line cap=round]
  \draw[thick] (18,294) -- (282,294);
  \node[anchor=north west, font=\scriptsize] at (20,298) {floor $f$};
  \begin{scope}[shift={(50,0)}, line width=1.1pt]
    \draw (100,66) -- (78,76) -- (70,118) -- (66,158);
    \draw (100,66) -- (122,76) -- (130,118) -- (134,158);
    \draw (100,66) -- (100,160);
    \draw (87,160) -- (113,160);
    \draw (87,160) -- (86,224) -- (85,290);
    \draw (113,160) -- (114,224) -- (115,290);
    \draw (75,292) -- (91,292);
    \draw (109,292) -- (125,292);
    \draw (100,38) circle (17);
    \foreach \p in {(100,66),(78,76),(122,76),(70,118),(130,118),(66,158),(134,158),(100,160),(87,160),(113,160),(86,224),(114,224),(85,290),(115,290)}
      \fill[orange!80!black] \p circle (1.5pt);
  \end{scope}
  \draw[-{Stealth[length=2mm]}, very thick] (262,270) -- (262,206) node[above right=-2pt, font=\small] {$\up$};
  \draw[-{Stealth[length=2mm]}, very thick] (262,270) -- (216,270) node[left, font=\small] {$\lft$};
  \foreach \n/\x/\y in {1/150/308, 2/186/224, 3/137/101, 4/214/140, 5/119/47}
    \node[circle, draw=orange!80!black, fill=white, thick, inner sep=0.8pt, font=\tiny\bfseries] at (\x,\y) {\n};
  \end{tikzpicture}
  \caption{The calibrated body frame: up $\up$, left $\lft$, floor $f$. Numbers mark where each move is measured: 1~Jump, 2~Squat, 3~Slip, 4~Punch, 5~Guard.}
  \label{fig:bodyframe}
\end{wrapfigure}
\textbf{Calibration and body frame.} At the start of a run, the player stands still, whole body in view, for 2~s (no joint moves more than 3~cm). This calibration defines a body frame (Fig.~\ref{fig:bodyframe}): the up direction $\up$ is the gravity vector of the phone's accelerometer (or, without it, the direction from the mid-ankles to the neck); the player's left $\lft = \mathrm{norm}(a - (a\cdot\up)\,\up)$, where $a$ is the mean vector from the right to the left shoulder, kept horizontal; the floor height $f$ is the mean height of the lowest foot point along $\up$, then slowly tracked while the player stands; and the neutral sideways lean $\lambda_0$ is recorded. Because InstantHMR outputs a metric 3D skeleton, every measure below is in metres or degrees, independent of the player's distance to the camera and of the camera's resolution.

\textbf{One measure, one threshold.} Each of the five moves is defined by a single measure computed on the 3D skeleton of each frame, and fires when it crosses a single threshold (Table~\ref{tab:gestures}); a few guards keep moves apart. Let $p_j$ denote joint positions in camera space and $\kappa = 180^\circ - \angle(\text{hip},\text{knee},\text{ankle})$ the flexion of a knee (0$^\circ$ for a straight leg).
\begin{itemize}
  \item \emph{Jump}: the lift $h - f$ of the lowest foot point $h = \min_{j\in\text{feet}} p_j\cdot\up$ exceeds 8~cm with straight knees ($\max\kappa < 40^\circ$).
  \item \emph{Squat}: both knees bend past $90^\circ$ ($\min(\kappa_L,\kappa_R) > 90^\circ$); bent knees within 800~ms after a jump are its landing and are ignored.
  \item \emph{Slip} (left or right): the neck leans sideways from the hips, $\lambda = (p_\text{neck} - \tfrac12(p_\text{hipL} + p_\text{hipR}))\cdot\lft - \lambda_0$, by more than 14~cm, the sign giving the side, while the feet stay planted (ankles moving less than 12~cm over 400~ms).
  \item \emph{Punch} (left or right arm): with $s$, $e$, $w$ the shoulder, elbow and wrist and $L = \|e-s\|+\|w-e\|$ the arm's length, the reach $r = \|w-s\|/L$ ($\approx 0.4$ with the fist at the chin, 1 with the arm straight) exceeds 0.6, having grown by at least 0.1 within 300~ms (an explosive extension), with the wrist not lowered ($(w-s)\cdot\up/L > -0.35$).
  \item \emph{Guard} (a state, not an event): with $c$ the midpoint of nose and neck, the distance from $c$ to the farther wrist, in shoulder widths, $g = \max_i \|w_i - c\| / \|s_L - s_R\|$, stays below 1.2; it is shown after 250~ms held.
\end{itemize}

\begin{table}[t]
\centering
\caption{The five moves: one measure on the metric 3D skeleton and one threshold each. Re-arm: the value the measure must return past before the move can fire again.}
\label{tab:gestures}
\small
\begin{tabular}{@{}>{\raggedright\arraybackslash}p{1.1cm} >{\raggedright\arraybackslash}p{6.2cm} >{\raggedright\arraybackslash}p{1.5cm} >{\raggedright\arraybackslash}p{2.2cm}@{}}
\toprule
\textbf{Move} & \textbf{Measure} & \textbf{Fires} & \textbf{Re-arms} \\
\midrule
Jump & lowest foot point above the floor $f$, knees straight & \mbox{$> 8$~cm} & $< 4$~cm \\
Squat & flexion of both knees & $> 90^\circ$ & $< 40^\circ$ \\
Slip & neck's sideways lean from the hips, along $\lft$, feet planted & \mbox{$> 14$~cm} & $< 7$~cm \\
Punch & reach: shoulder--wrist distance / arm length, grown by $\geq 0.1$ in 0.3~s & $> 0.6$ & $< 0.55$, after 0.3~s \\
Guard & farther fist to chin, in shoulder widths & $< 1.2$ & wrist lowered \\
\bottomrule
\end{tabular}
\end{table}

\textbf{Re-arming instead of smoothing.} We apply no temporal filter to the skeleton, as smoothing would delay every move. Instead, each move uses hysteresis and refractory periods: after firing, it re-arms only once its measure has come back past a lower value (Table~\ref{tab:gestures}), so a single physical movement fires exactly once, and noise around a threshold does not produce double triggers.

\textbf{Personal thresholds.} The default thresholds suit most players, but not all: some cannot jump 8~cm or squat past 90$^\circ$. During the tutorial (Sect.~\ref{sec:game}), the game records the peak of the measure on each attempt. After three near misses (peaks reaching at least 70\% of the way from rest to the threshold), the threshold is lowered to $\theta' = \max(\theta_\text{floor},\ m_\text{rest} + 0.95\,(m_\text{best} - m_\text{rest}))$, where $m_\text{rest}$ is the measure at rest and $m_\text{best}$ the best attempt. Thresholds are only ever lowered, and never below safety floors (jump 8~cm, squat 40$^\circ$, slip 11~cm, punch reach 0.59), the lowest values at which the moves stayed distinct from each other and from ordinary movement on our labelled recordings. Personal thresholds are saved in the player's profile.

\subsection{Gameplay}
\label{sec:game}

The player controls the fighter Axel. Each move maps to one action: a right punch triggers a punch combo and a left punch a kick; a slip dashes toward the nearest enemy on that side, invulnerable (a punch within 500~ms after the dash is a counter, dealing double damage); a squat triggers an uppercut on both sides and a jump a leap onto the nearest enemy (with 2~s and 2.5~s cooldowns); a held guard halves incoming damage. Enemy attacks start with a 600~ms wind-up, leaving time to slip or guard, and the fighter advances automatically between waves, so the player never navigates.

\begin{figure}[t]
  \centering
  \includegraphics[width=0.329\textwidth,alt={Tutorial lesson for the right punch: a panel at the top left reads ``Right punch: punch straight at the camera with your right arm'', a coach stick figure labelled COPY ME shows the move, and the fighter Axel punches an enemy. The player's camera view with the 3D skeleton is in the top-right corner.}]{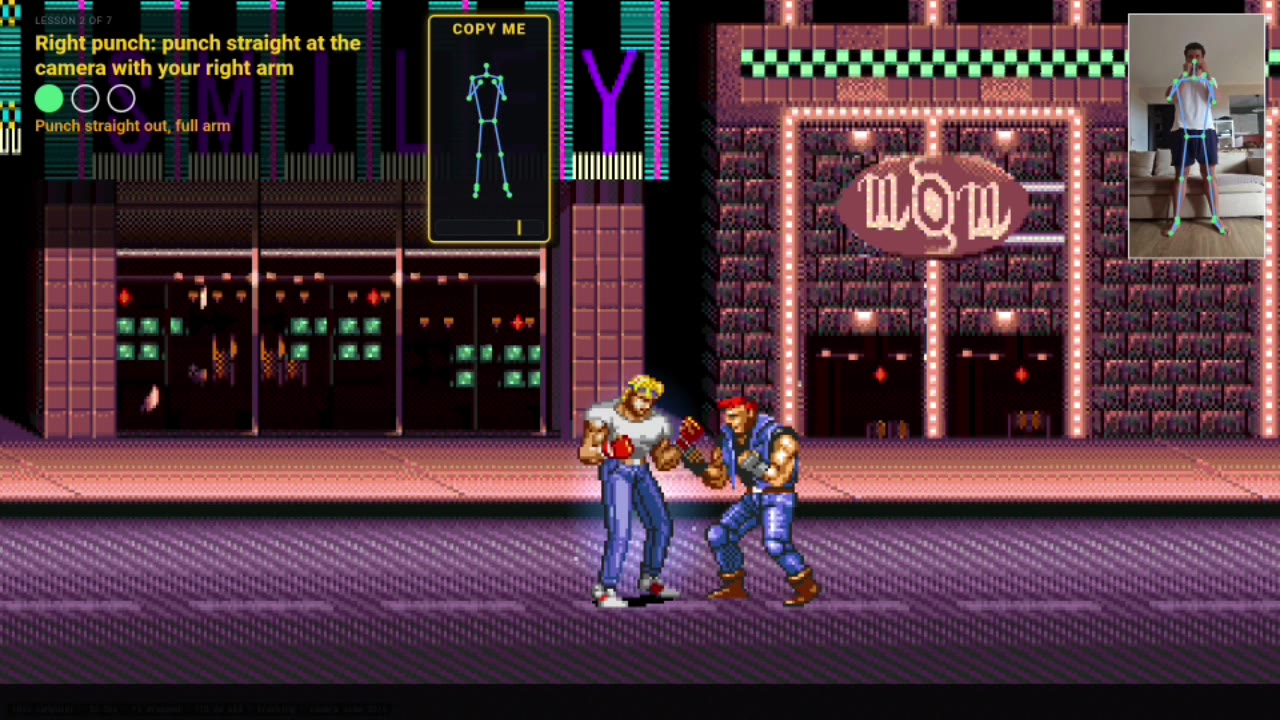}\hfill
  \includegraphics[width=0.329\textwidth,alt={Street wave: Axel fights the large enemy Big Ben and two other enemies; two buttons at the bottom show the squat and jump specials. The player's camera view with the 3D skeleton is in the top-right corner.}]{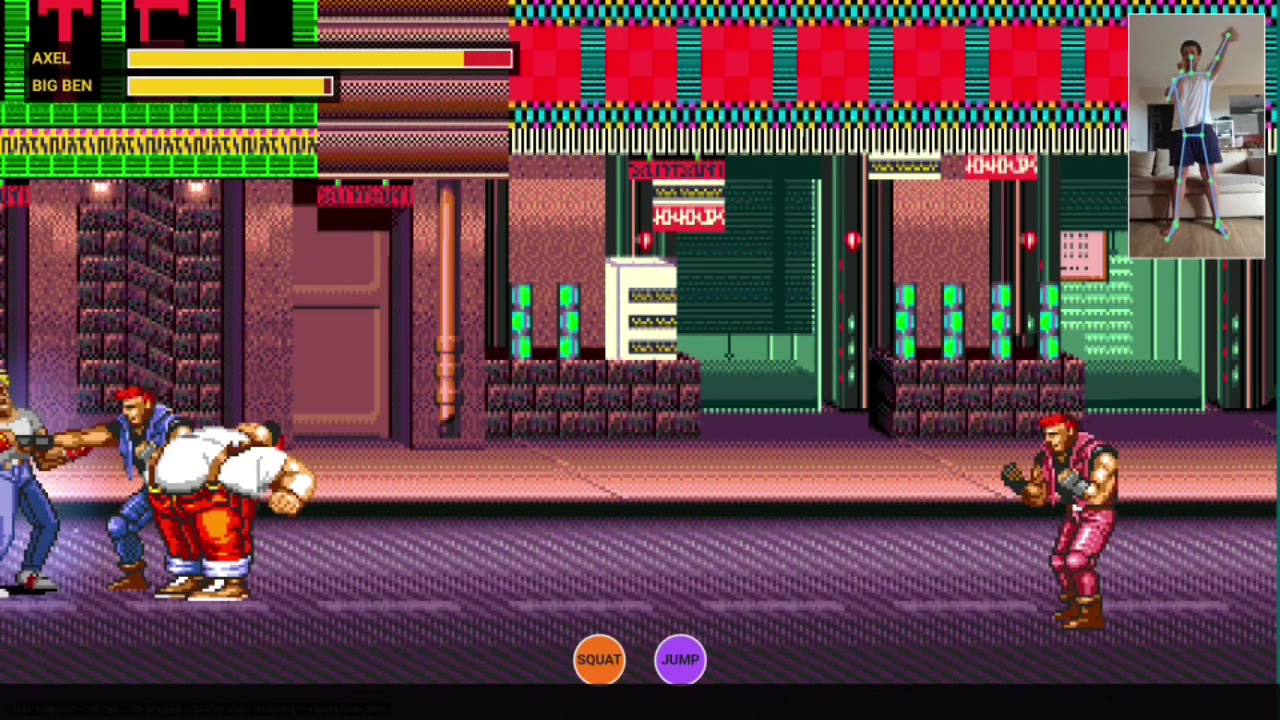}\hfill
  \includegraphics[width=0.329\textwidth,alt={Rhythm stage, showdown against Curls: three fireballs fly toward Axel, each with a punch cue (R, L, R) and a green ring closing on it; a 15 COMBO counter is shown at the top.}]{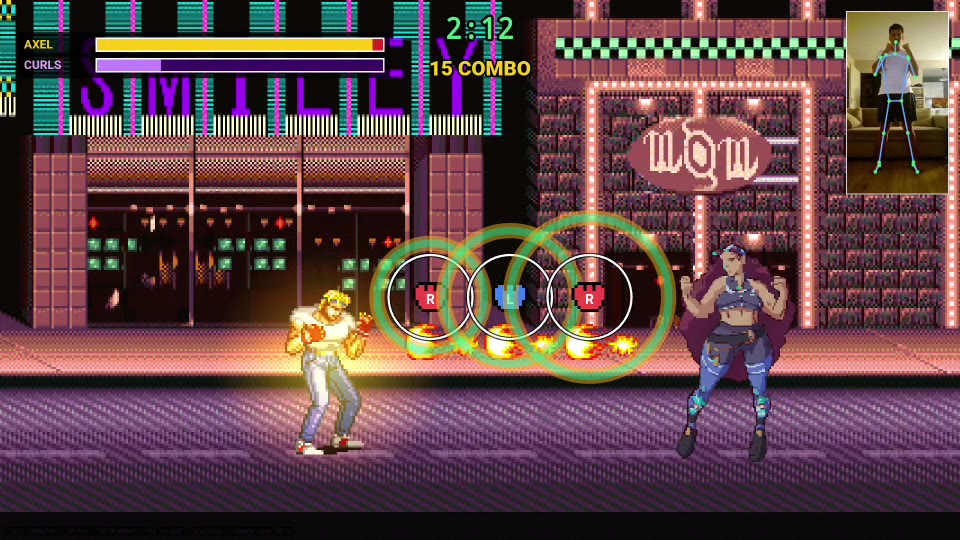}
  \caption{Left: a tutorial lesson, where the coach shows the move (here, the right punch) and the player repeats it. Middle: a street wave against the enemy Big Ben, fought freely. Right: the rhythm stage, a volley from the boss Curls, three fireballs, each a punch cue (right, left, right) whose ring closes on the kick drum (music: ``Shiny Tech'' by Kevin MacLeod, incompetech.com, CC BY 4.0). The player's camera view, with the 3D skeleton, is in the top-right corner.}
  \label{fig:screens}
\end{figure}

A run lasts about 3.5 minutes. After calibration and the choice of a profile, a tutorial teaches the moves one by one: for each lesson, a coach figure shows the move (``COPY ME'') and the player performs it three times, while personal thresholds are adjusted (Fig.~\ref{fig:screens}, left). Then come two free street waves, where enemies attack at their own pace and the player reacts (Fig.~\ref{fig:screens}, middle), and finally the rhythm stage.

\subsection{Playing on the Beat}
\label{sec:rhythm}

\textbf{Cues on the kick drum.} In the rhythm stage, enemies attack on the music, and each attack is announced by a cue: the icon of the expected move over the enemy, with a ring closing on it that meets it exactly on the beat (Fig.~\ref{fig:screens}, right). We use discrete cues rather than continuous, cyclic movement such as jumping or squatting on every beat: discrete cues fit into level design (an enemy's attack, a volley of fireballs, a boss's sequence) and do not exhaust the player, whereas a minute of continuous jumping does. Cues are placed on the beats that a player hears best, those with a strong kick drum, so the song must be fast and have a kick on almost every beat. We measured the beat grid and kick strength of 26 royalty-free tracks offline, with a dynamic-programming beat tracker~\cite{Ellis2007} and the onset strength in the bass band ($<200$~Hz), and chose ``Shiny Tech''~\cite{ShinyTech}: 138~BPM, a constant tempo, and a kick on 90\% of its beats, which gives 119 beats on which cues can be placed. The stage alternates punch cues, a thrower's fireballs (one punch cue each) and a showdown against the boss Curls, whose turns chain several cues into sequences. Cues appear one to five beats before their beat, depending on the attack. The song is decoded once and started on the Web Audio clock, and each beat is mapped to the page's clock from the audio output timestamp, so the beats reach the player where the game placed its cues.

\textbf{Judgement on capture time.} A move is recognised 100--250~ms after the body performed it: the pose pipeline, the wait for the next frame and the recognition itself add delays that vary with the move, the phone's temperature and the setup. Judging the recognition time would make players who hit the beat read as late, and the rhythm would feel different on every setup. Instead, each judgement compares the beat $b$ with the camera capture time $t_\text{cap}$ of the frame that completed the move, which the phone stamps from the camera sensor and which the game converts to the page's clock:
\begin{equation}
  e = t_\text{cap} + a_g - b - d,
\end{equation}
where $a_g$ is a per-move adjustment for the point of the movement at which it is recognised (a punch is recognised about 120~ms before full extension, which the player aims at the beat, so $a_g = +120$~ms; a jump is recognised about 50~ms after takeoff, $a_g = -50$~ms), and $d$ is a delay learned for each player: the median error of their first eight correct answers within $\pm 400$~ms. This delay absorbs, without any calibration screen, the latency of the sound output and of the TV, as well as the player's own habit of anticipating or trailing the beat. A cue is judged Perfect if $|e| \le 130$~ms, Good if $|e| \le 260$~ms (shown as EARLY or LATE), and Miss otherwise or for the wrong move; slips, squats and jumps, which are slower whole-body movements, get wider windows ($\pm 180$ and $\pm 340$~ms). At the end of the run, a rank summarises the share of cues on the beat, $100\,(N_\text{Perfect} + N_\text{Good}/2)/N$ (S from 90, A from 75, B from 50).

\section{Design Rationale}
\label{sec:rationale}

The choice of a side-scrolling brawler is semantically motivated. Shadow boxing, a primary modality we target, shares its core vocabulary with the genre: punches, dodges, guard and rhythmic aggression. Mapping physical strikes to avatar attacks reduces the cognitive distance between movement and on-screen consequence, and the avatar acts as a corporeal proxy whose combat provides embodied feedback for the sensorimotor synchronization loop~\cite{Repp2013}. Short runs of a few minutes, alternating reactive fights and rhythmic sequences, fit the exercise snack paradigm~\cite{Fang2026,Stamatakis2022}.

Recognising moves on a metric 3D skeleton rather than on 2D keypoints lets each move be described by a measure a coach would use (centimetres of lift, degrees of knee flexion, the reach of an arm), independent of the player's distance or orientation to the camera, and the phone's gravity sensor keeps the body frame vertical even when the phone is tilted. One measure and one threshold per move keep the recognition easy to explain, to personalise, and to tune from labelled recordings. Finally, running everything on the phone removes the last piece of dedicated hardware: a phone and a USB-C to HDMI cable, which many households already own, are enough. Keeping the game in web technologies lets the same game run in a desktop browser during development, and provides the Web Audio clock needed to schedule the music precisely.

\section{Limitations and Future Work}
\label{sec:future}

This work in progress has several limitations. No user study has been conducted yet: the thresholds and windows were tuned through informal play by the authors, and the safety floors come from one player's labelled recordings. Informal play feels reactive, but the latencies are not all compensated: feedback (hit, sound and judgement) necessarily comes after the recognition, cues are drawn without compensating the display and TV latency, the learned delay is learned on punches only, and the mapping between the audio and page clocks is sampled once at the start of the song. The game has one song and one level, the per-move adjustments are starting values, and the sprites are placeholders from the original game.

\textbf{Planned evaluation.} A user study will assess the system along three dimensions: (1) the reliability of the move recognition, by comparing detected moves against manually annotated video; (2) perceived exertion and engagement, contrasting the rhythm stage with the free street waves and with a version without music; and (3) rhythmic accuracy, using the capture-time errors already logged for every judgement. It will be a within-subject exploratory study with healthy adults, combining telemetry with validated self-report instruments.

\textbf{Future developments.} We plan to compensate the remaining latencies (display, TV, audio clock drift), to add songs, tempos and levels, to introduce short bursts of cyclic movement (e.g., a few seconds of jumping on the beat) between discrete cues, and to create original art.

\section{Conclusion}

We have presented a rhythmic beat 'em up exergame that turns a smartphone and an HDMI cable into an exergame console. An off-the-shelf real-time human mesh recovery model running on the phone~\cite{InstantHMR} provides a metric 3D skeleton, on which five shadow-boxing moves are recognised with one measure and one threshold each, personalised during the tutorial. A rhythm stage places discrete cues on the kick drum of a fast song and judges them on the movement's capture time. Informal play suggests that full-body, beat-driven action is achievable on commodity hardware; the planned user study will test recognition reliability, engagement, perceived exertion~\cite{Karageorghis2012}, and rhythmic accuracy.

\subsubsection*{Acknowledgments.} This work was supported by the Eurostars programme (Eureka / European Union), project 9611 NeuroComp (NeuroCompanion for Parkinson disease management).

\bibliographystyle{splncs04}
\bibliography{references_icec2026}

\begin{thebibliography}{10}
\providecommand{\url}[1]{\texttt{#1}}
\providecommand{\urlprefix}{URL }
\providecommand{\doi}[1]{https://doi.org/#1}

\bibitem{Ballmann2021}
Ballmann, C.G.: The influence of music preference on exercise responses and performance: {A} review. Journal of Functional Morphology and Kinesiology  \textbf{6}(2), ~33 (2021). \doi{10.3390/jfmk6020033}

\bibitem{Bayd2025}
Bayd, H., Guyot, P., Bardy, B., Slangen, P.: Influence of rhythm features on beat/movement synchronization using a low-cost vision system. Frontiers in Computer Science  \textbf{7},  1595939 (2025). \doi{10.3389/fcomp.2025.1595939}

\bibitem{Begel2017}
B{\'e}gel, V., Di~Loreto, I., Seilles, A., Dalla~Bella, S.: Music games: Potential application and considerations for rhythmic training. Frontiers in Human Neuroscience  \textbf{11}, ~273 (2017). \doi{10.3389/fnhum.2017.00273}

\bibitem{BraunJanzen2021}
Braun~Janzen, T., Koshimori, Y., Richard, N.M., Thaut, M.H.: Rhythm and music-based interventions in motor rehabilitation: Current evidence and future perspectives. Frontiers in Human Neuroscience  \textbf{15},  789467 (2021). \doi{10.3389/fnhum.2021.789467}

\bibitem{openpose}
Cao, Z., Martinez, G., Simon, T., Wei, S.E., Sheikh, Y.: {OpenPose}: Realtime multi-person 2d pose estimation using part affinity fields. IEEE Transactions on Pattern Analysis and Machine Intelligence  \textbf{PP}, ~1--1 (2019). \doi{10.1109/TPAMI.2019.2929257}

\bibitem{CochenDeCock2021}
Cochen De~Cock, V., Dotov, D., Damm, L., Lacombe, S., Ihalainen, P., Picot, M.C., Galtier, F., Lebrun, C., Giordano, A., Driss, V., Geny, C., Garzo, A., Hernandez, E., Van~Dyck, E., Leman, M., Villing, R., Bardy, B.G., Dalla~Bella, S.: {BeatWalk}: Personalized music-based gait rehabilitation in {Parkinson's} disease. Frontiers in Psychology  \textbf{12},  655121 (2021). \doi{10.3389/fpsyg.2021.655121}

\bibitem{Craig2024}
Craig, T.V., Rhodes, R.E., Sui, W.: Examining and comparing the energy expenditure of two modes of a virtual reality fitness game ({Supernatural}): indirect calorimetry study. JMIR Serious Games  \textbf{12},  e53999 (2024). \doi{10.2196/53999}

\bibitem{Begel2018}
Dauvergne, C., B{\'e}gel, V., Geny, C., Puyjarinet, F., Laffont, I., Dalla~Bella, S.: Home-based training of rhythmic skills with a serious game in {Parkinson's} disease: Usability and acceptability. Annals of Physical and Rehabilitation Medicine  \textbf{61}(6),  380--385 (2018). \doi{10.1016/j.rehab.2018.08.002}

\bibitem{Ellis2007}
Ellis, D.P.W.: Beat tracking by dynamic programming. Journal of New Music Research  \textbf{36}(1),  51--60 (2007). \doi{10.1080/09298210701653344}

\bibitem{Fang2026}
Fang, Y., Li, H., Dong, P., Wan, F.: Micro-exercise breaks every hour: a feasible strategy to improve metabolic health in sedentary office workers. BMC Public Health  \textbf{26} (2026). \doi{10.1186/s12889-026-26484-4}

\bibitem{Ferguson2025MHR}
Ferguson, A., Osman, A.A.A., Bescos, B., et~al.: {MHR}: {Momentum Human Rig}. arXiv:2511.15586 (2025)

\bibitem{Grahn2009}
Grahn, J.A., Brett, M.: Impairment of beat-based rhythm discrimination in {Parkinson's} disease. Cortex  \textbf{45}(1),  54--61 (2009). \doi{10.1016/j.cortex.2008.01.005}

\bibitem{Grech2024}
Grech, E., Briguglio, M., Said, E.: A field experiment on gamification of physical activity: effects on motivation and steps. International Journal of Human-Computer Studies  \textbf{184},  103205 (2024). \doi{10.1016/j.ijhcs.2023.103205}

\bibitem{Kanazawa2018}
Kanazawa, A., Black, M.J., Jacobs, D.W., Malik, J.: End-to-end recovery of human shape and pose. In: Proceedings of the IEEE Conference on Computer Vision and Pattern Recognition (CVPR) (2018). \doi{10.1109/CVPR.2018.00744}

\bibitem{Karageorghis2012}
Karageorghis, C.I., Priest, D.L.: Music in the exercise domain: a review and synthesis ({Part I}). International Review of Sport and Exercise Psychology  \textbf{5}(1),  44--66 (2012). \doi{10.1080/1750984X.2011.631026}

\bibitem{Karaosmanoglu2024}
Karaosmanoglu, S., Cmentowski, S., Nacke, L.E., Steinicke, F.: Born to run, programmed to play: mapping the extended reality exergames landscape. In: Proceedings of the CHI Conference on Human Factors in Computing Systems (CHI '24). ACM (2024). \doi{10.1145/3613904.3642124}

\bibitem{Large1999}
Large, E.W., Jones, M.R.: The dynamics of attending: How people track time-varying events. Psychological Review  \textbf{106}(1),  119--159 (1999). \doi{10.1037/0033-295X.106.1.119}

\bibitem{mediapipe}
Lugaresi, C., Tang, J., Nash, H., McClanahan, C., Uboweja, E., Hays, M., Zhang, F., Chang, C.L., Yong, M., Lee, J., Chang, W.T., Hua, W., Georg, M., Grundmann, M.: {MediaPipe}: A framework for building perception pipelines. arXiv preprint arXiv:1906.08172  (2019)

\bibitem{ShinyTech}
MacLeod, K.: Shiny tech. \url{https://incompetech.com}, licensed under Creative Commons: By Attribution 4.0

\bibitem{Monroe2023}
Monroe, C., Cai, B., Edney, S., Jake-Schoffman, D., Brazendale, K., Bucko, A., Armstrong, B., Yang, C.H., Turner-McGrievy, G.: Harnessing technology and gamification to increase adult physical activity: a cluster randomized controlled trial of the {Columbia Moves} pilot. International Journal of Behavioral Nutrition and Physical Activity  \textbf{20} (2023). \doi{10.1186/s12966-023-01530-1}

\bibitem{InstantHMR}
{NaturalPad}: {InstantHMR}: real-time human mesh recovery on edge devices. \url{https://github.com/Naturalpad-studio/InstantHMR} (2026), accessed October 2026

\bibitem{OhYang2010}
Oh, Y., Yang, S.: Defining exergames and exergaming. In: Proceedings of Meaningful Play. pp. 1--17. MSU Serious Games Program, East Lansing, MI (2010)

\bibitem{Peng2011}
Peng, W., Lin, J.H., Crouse, J.: Is playing exergames really exercising? {A} meta-analysis of energy expenditure in active video games. Cyberpsychology, Behavior, and Social Networking  \textbf{14}(11),  681--688 (2011). \doi{10.1089/cyber.2010.0578}

\bibitem{Phosanarack2025}
Phosanarack, M., Wallard, L., Lepreux, S., Kolski, C., Avril, E.: Smartphone exergames with real-time markerless motion capture: challenges and trade-offs. CHI '25 Workshop, arXiv:2507.06669 (2025)

\bibitem{Pranjic2024}
Pranji{\'c}, M., Braun~Janzen, T., Vuk{\v{s}}i{\'c}, N., Thaut, M.: From sound to movement: Mapping the neural mechanisms of auditory--motor entrainment and synchronization. Brain Sciences  \textbf{14}(11), ~1063 (2024). \doi{10.3390/brainsci14111063}

\bibitem{Puyjarinet2022}
Puyjarinet, F., B{\'e}gel, V., Geny, C., Driss, V., Cuartero, M.C., De~Cock, V.C., Pinto, S., Dalla~Bella, S.: At-home training with a rhythmic video game for improving orofacial, manual, and gait abilities in {Parkinson's} disease: A pilot study. Frontiers in Neuroscience  \textbf{16},  874032 (2022). \doi{10.3389/fnins.2022.874032}

\bibitem{NanoDet}
{RangiLyu}: {NanoDet}: super fast and lightweight anchor-free object detection model. \url{https://github.com/RangiLyu/nanodet} (2021)

\bibitem{CityOfRage}
{raytomely}: {City of Rage}: a {Streets of Rage 2} tribute in {Pygame}. \url{https://raytomely.itch.io/city-of-rage}

\bibitem{Repp2005}
Repp, B.H.: Sensorimotor synchronization: A review of the tapping literature. Psychonomic Bulletin \& Review  \textbf{12}(6),  969--992 (2005). \doi{10.3758/BF03206433}

\bibitem{Repp2013}
Repp, B.H., Su, Y.H.: Sensorimotor synchronization: A review of recent research (2006--2012). Psychonomic Bulletin \& Review  \textbf{20}(3),  403--452 (2013). \doi{10.3758/s13423-012-0371-2}

\bibitem{Richard2025}
Richard, P., Zaoui, Y., Adjel, M., B{\'e}gel, V., Rodriguez, N.: {AutonoMoov}: an embedded physio-gaming platform with multi-level and rhythmic motor interaction. In: IHM 2025 - 36e Conf{\'e}rence Internationale Francophone sur l'Interaction Humain-Machine (2025)

\bibitem{Sarandi2024}
S{\'a}r{\'a}ndi, I., Pons-Moll, G.: Neural localizer fields for continuous 3d human pose and shape estimation. In: Advances in Neural Information Processing Systems (NeurIPS) (2024). \doi{10.52202/079017-4444}

\bibitem{Schone2025}
Sch{\"o}ne, C., Sauter, M., Back{\'e}, E.M., Prigge, M., Brendler, C., Hegewald, J.: The impact of working from home on sedentary behaviour and physical activity compared to onsite work in the working population: a systematic review and meta-analysis. BMC Public Health  \textbf{25} (2025). \doi{10.1186/s12889-025-24960-x}

\bibitem{SOR2}
{Sega}: Streets of rage 2. Video game, Mega Drive/Genesis (1992)

\bibitem{Stamatakis2022}
Stamatakis, E., Ahmadi, M.N., Gill, J.M.R., et~al.: Association of wearable device-measured vigorous intermittent lifestyle physical activity with mortality. Nature Medicine  \textbf{28}(12),  2521--2529 (2022). \doi{10.1038/s41591-022-02100-x}

\bibitem{Strain2024}
Strain, T., Flaxman, S., Guthold, R., et~al.: National, regional, and global trends in insufficient physical activity among adults from 2000 to 2022. The Lancet Global Health  \textbf{12},  e1232--e1243 (2024). \doi{10.1016/S2214-109X(24)00150-5}

\bibitem{Terry2020}
Terry, P.C., Karageorghis, C.I., Curran, M.L., Martin, O.V., Parsons-Smith, R.L.: Effects of music in exercise and sport: {A} meta-analytic review. Psychological Bulletin  \textbf{146}(2),  91--117 (2020). \doi{10.1037/bul0000216}

\bibitem{Thaut2019}
Thaut, M.H., Rice, R.R., Braun~Janzen, T., Hurt-Thaut, C.P., McIntosh, G.C.: Rhythmic auditory stimulation for reduction of falls in {Parkinson's} disease: a randomized controlled study. Clinical Rehabilitation  \textbf{33}(1),  34--43 (2019). \doi{10.1177/0269215518788615}

\bibitem{Wang2022}
Wang, L., Peng, J., Ou-Yang, J., Gan, L., Zeng, S., Wang, H.Y., Zuo, G.C., Qiu, L.: Effects of rhythmic auditory stimulation on gait and motor function in {Parkinson's} disease: {A} systematic review and meta-analysis. Frontiers in Neurology  \textbf{13},  818559 (2022). \doi{10.3389/fneur.2022.818559}

\bibitem{Whitton2023}
Whitton, S.A., Jiang, F.: Sensorimotor synchronization with visual, auditory, and tactile modalities. Psychological Research  \textbf{87}(7),  2204--2217 (2023). \doi{10.1007/s00426-023-01801-3}

\bibitem{Yang2026SAM3DBody}
Yang, X., Kukreja, D., Pinkus, D., et~al.: {SAM 3D Body}: robust full-body human mesh recovery. arXiv:2602.15989 (2026)

\bibitem{Zhao2025}
Zhao, J., Zhu, D., Wang, Y., et~al.: Development and evaluation of a monocular camera--based mobile exergame for at-home intervention in individuals at high risk of type 2 diabetes: randomized controlled trial. JMIR Serious Games  \textbf{13},  e75823 (2025). \doi{10.2196/75823}

\end{thebibliography}

\end{document}